\documentclass{article}

\usepackage[preprint]{neurips_2025}
\usepackage[utf8]{inputenc} % allow utf-8 input 
\usepackage[T1]{fontenc}    % use 8-bit T1 fonts
\usepackage{hyperref}       % hyperlinks
\usepackage{url}            % simple URL typesetting
\usepackage{booktabs}       % professional-quality tables
\usepackage{amsfonts}       % blackboard math symbols
\usepackage{nicefrac}       % compact symbols for 1/2, etc.
\usepackage{microtype}      % microtypography
\usepackage{xcolor}         % colors
\usepackage{graphicx}
\usepackage{wrapfig}  % 在导言区添加

\usepackage{authblk}
\usepackage{subfiles}

\usepackage{algorithm}
\usepackage[noend]{algpseudocode}

\usepackage{color}

\usepackage{hyperref} 
\hypersetup{
    colorlinks=true,
    linkcolor=red,
    urlcolor=blue,
    linktoc=all,
    citecolor=cyan
}

\usepackage{enumitem}
\usepackage{lipsum}
\setlist[itemize]{leftmargin=*}
 
\definecolor{BgWhite}{rgb}{1,1,1} 
\definecolor{Gray}{rgb}{0.5,0.5,0.5} 
\definecolor{mygray}{gray}{0.6}

\usepackage{multirow}
\usepackage[utf8]{inputenc} % allow utf-8 input 
\usepackage{amsmath}

\usepackage{xcolor}

\title{TopGen: Learning Structural Layouts and Cross-Fields for Quadrilateral Mesh Generation}
\author{
Yuguang Chen$^{1,2}$, Xinhai Liu$^{2}$, Xiangyu Zhu$^{2}$, Yiling Zhu$^{2}$,\\ Zhuo Chen$^2$, Dongyu Zhang$^{1,\ddag}$, Chunchao Guo$^{2,\ddag}$ \\ \vspace{0.3cm}
$^1$SYSU, $^2$ Tencent Hunyuan \\ \vspace{0.3cm}
}

\begin{document}

\maketitle
\begin{figure}[h]
    \centering
    \includegraphics[width=\textwidth]{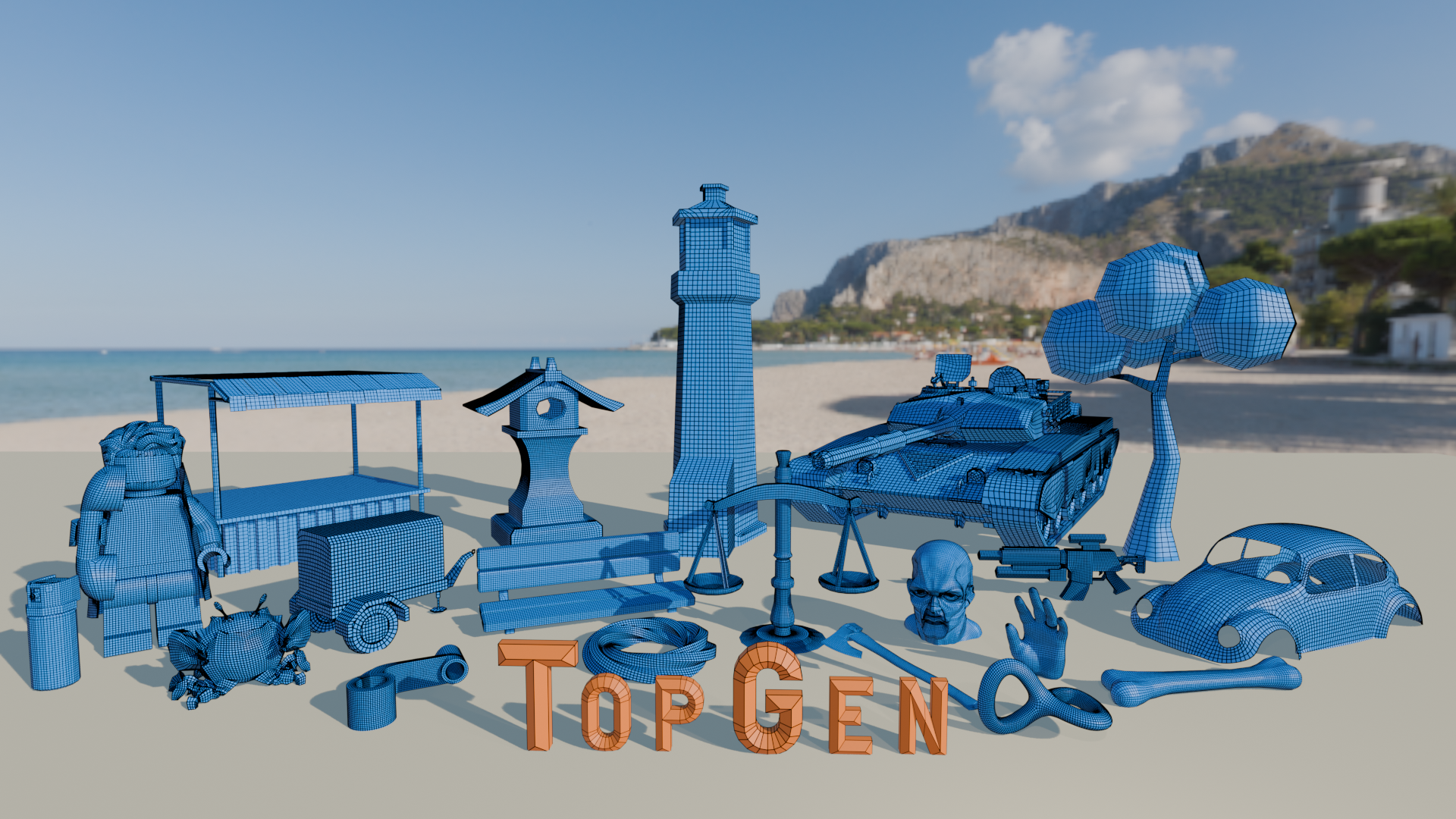}
    \caption{TopGen achieves high-quality quadrilateral remeshing across diverse geometries, ranging from organic to mechanical models and AI-generated meshes.}
    \label{fig:teaser}
\end{figure}
\begin{abstract}
High-quality quadrilateral mesh generation is a fundamental challenge in computer graphics. Traditional optimization-based methods are often constrained by the topological quality of input meshes and suffer from severe efficiency bottlenecks, frequently becoming computationally prohibitive when handling high-resolution models. While emerging learning-based approaches offer greater flexibility, they primarily focus on cross-field prediction, often resulting in the loss of critical structural layouts and a lack of editability. In this paper, we propose TopGen, a robust and efficient learning-based framework that mimics professional manual modeling workflows by simultaneously predicting structural layouts and cross-fields. By processing input triangular meshes through point cloud sampling and a shape encoder, TopGen is inherently robust to non-manifold geometries and low-quality initial topologies. We introduce a dual-query decoder—utilizing edge-based and face-based sampling points as queries—to perform structural line classification and cross-field regression in parallel. This integrated approach explicitly extracts the geometric "skeleton" while concurrently capturing orientation fields. Such synergy ensures the preservation of geometric integrity and provides an intuitive, editable foundation for subsequent quadrilateral remeshing. To support this framework, we also introduce a large-scale, comprehensive quadrilateral mesh dataset, TopGen-220K, featuring high-quality paired data comprising raw triangular meshes, structural layouts, cross-fields, and their corresponding quad-meshes. Experimental results demonstrate that TopGen significantly outperforms existing state-of-the-art methods in both geometric fidelity and topological edge flow rationality.
\end{abstract}

\section{Introduction}
\label{sec:intro}

Quadrilateral meshes (or Quad Meshes) are a fundamental 3D representation in computer graphics and geometric modeling. As the industry standard for character modeling and high-fidelity surface representation, they play a critical role in animation, industrial design, and digital content creation. With the rapid advancement of 3D generative techniques~\cite{siddiqui2024meshgpt, zhao2025hunyuan3d,xiang2025native}, high-resolution models with rich details can now be synthesized automatically. However, the resulting topologies are often fragmented and chaotic, rendering them incompatible with professional game and production pipelines. Thus, a fast and high-precision method for quadrilateral mesh generation is of significant research and practical value.

In industrial production, high-quality quad meshes are typically authored by skilled artists following a "layout-first, generation-second" workflow. They first meticulously mark structural lines on the model surface to establish an organized structural layout, then utilize retopology tools (e.g., ZBrush ZRemesher) to convert dense triangular meshes into regular quadrilateral grids. While this manual process ensures peak topological rationality and faithful alignment with design intent, it remains an extremely labor-intensive and time-consuming task. Consequently, this bottleneck hinders its integration into large-scale production pipelines and fails to keep pace with the rapid surge in AI-generated 3D content.

Historically, the academic community has focused on optimization-based automated remeshing~\cite{jakob2015instant,bommes2009mixed,huang2018quadriflow,pietroni2021reliable, tao2025learning}. These algorithms rely on cross-field guidance and global parametrization to facilitate the triangle-to-quad transition~\cite{sumner2004deformation}. They typically construct smooth cross-fields with sparse constraints to determine singularity placements, followed by surface cutting~\cite{campen2017partitioning} and parametrization. However, the high complexity of global optimization leads to severe efficiency bottlenecks, particularly for high-resolution meshes. Furthermore, these methods demand high-quality manifold input and struggle with complex features, often resulting in irrational singularity distributions, distorted edge flows, and a lack of high-level semantic awareness.

Recently, learning-based quad remeshing has attracted significant attention. These methods~\cite{dong2025neurcross,dong2025crossgen,liu2025neuframeq} focus on frame-field or cross-field prediction to overcome the limitations of Optimization-based approaches, such as heavy reliance on local geometry and low optimization efficiency. By leveraging auto-encoder architecture~\cite{dong2025crossgen} or diffusion models~\cite{liu2025neuframeq}, these techniques learn the mapping between geometric features and cross-fields from large-scale datasets, subsequently using the predicted fields to guide quad-mesh generation. However, cross-fields act only as "soft constraints" and inherently struggle to resolve geometric discontinuities, such as sharp edges and corners. This limitation often results in jagged boundaries and a loss of critical geometric detail, making it difficult to produce truly game-ready quad meshes that meet professional industry standards.

Therefore, our work is motivated by two key challenges. First, traditional optimization-based methods suffer from low computational efficiency and struggle to handle complex geometries or integrate high-level semantic information. Second, existing learning-based approaches rely exclusively on cross-fields as guidance; this dependency often leads to compromised geometric fidelity, loss of sharp details, and a lack of intuitive editability in the generated quad-meshes.

In this paper, we propose TopGen, a novel learning-based framework that generates quadrilateral meshes by jointly predicting structural layouts and cross-fields. Mimicking the professional "layout-first, generation-second" workflow, TopGen utilizes a unified Encoder-Decoder architecture to predict structural lines and cross-fields in parallel. In this framework, the structural lines act as hard constraints to ensure global structural integrity, while the cross-fields serve as soft constraints to optimize internal edge flow.

Specifically, we employ a point cloud sampling strategy coupled with the Sharp Edge Sampling technique~\cite{chen2025dora}. It samples points both along sharp features and uniformly across the surface of the input mesh to preserve fine-grained geometric details. These points are processed by a powerful pre-trained shape encoder that compresses geometric features into a continuous latent space. Then, we introduce a Dual Query Decoder. We sample query points at the midpoints of all edges and the centroids of all faces, termed edge queries and face queries, respectively. We then model structural line prediction as an edge-based classification task and cross-field prediction as a face-based regression task. These queries aggregate local topological information with neighborhood points through attention mechanisms and then extract structural lines and cross-fields from the global latent space via a cross-attention mechanism.

To facilitate model training, we construct TopGen-220K, a large-scale quadrilateral mesh dataset. This dataset provides a comprehensive suite of paired data, including raw triangular meshes, high-quality structural layouts, cross-fields, and their corresponding quad-meshes. Once trained, our framework demonstrates exceptional robustness in processing meshes with diverse topological qualities and surface attributes. Notably, it can predict high-quality structural lines and cross-fields in under one second. Guided by these structural and field-based constraints, our method further synthesizes high-fidelity quadrilateral meshes that align with rigorous professional artistic standards.

In summary, our primary contributions are as follows:
\begin{itemize}
\item We present TopGen, the first learning-based quadrilateral mesh generation framework to jointly predict structural layouts and cross-fields, ensuring both the integrity of external geometric boundaries and the regularity of internal edge flows.
\item We introduce a novel Dual-Query Decoder that leverages the synergy between structure lines and cross-fields to perform parallel edge classification and field regression.
\item We contribute a significant large-scale dataset providing paired triangular meshes, high-quality structure lines, cross-fields, and high-fidelity quadrilateral meshes, filling a critical gap in the field.
\end{itemize}

\section{Related work}
\label{sec:related_work}
\subsection{Optimization-based Quad Mesh Generation Method}
Traditional optimization-based quadrilateral remeshing~\cite{jakob2015instant,bommes2009mixed,huang2018quadriflow,huang2018quadriflow} has evolved through diverse frameworks centered on direction field design, surface segmentation, and the trade-off between global and local optimization. Based on their primary guidance mechanisms, these methods are generally categorized into cross-field-guided and feature-line-guided approaches.

Among cross-field-guided methods, Instant-Meshes~\cite{jakob2015instant} employs a local optimization paradigm. By utilizing a unified local smoothing operator based on $N$-RoSy direction fields and position fields, it achieves linear time complexity and parameter-independent alignment with surface features, enabling the efficient processing of large-scale meshes and point clouds. MIQ~\cite{bommes2009mixed} proposes a pipeline that first identifies critical bending directions to construct smooth cross-fields with automatically localized singularities, followed by mixed-integer linear programming to facilitate mesh cutting, flattening, and tiling. QuadriFlow~\cite{huang2018quadriflow} further improves upon Instant-Meshes by integrating a hybrid strategy—combining min-cost network flow with local Boolean satisfiability solvers—to significantly reduce the number of singularities.

In the realm of feature-line-guided methods, QuadWild~\cite{pietroni2021reliable} introduces a feature-driven patch partition scheme. By optimizing the input mesh and constructing feature-aligned cross-fields, it generates a polygonal layout bounded by feature lines, resulting in pure quad meshes with superior feature preservation. FSCQ~\cite{liang2025field} optimizes patch generation by prioritizing path tracing in regions with high field smoothness, effectively minimizing distortion caused by singularities. Furthermore, Frame Fields~\cite{panozzo2014frame} extend direction fields to anisotropic and non-orthogonal domains, providing flexible control over edge orientations. FlowRep~\cite{gori2017flowrep} extracts dominant streamlines from curvature-aligned meshes and combines them with sharp feature curves to construct a compact 3D curve network. This network serves as a structural layout to guide quad mesh generation.

Despite their rigor, these optimization-based methods suffer from high computational costs and potential convergence failures on complex models. Furthermore, they are sensitive to input quality and lack semantic awareness, often producing edge flow that deviate from professional artistic standards.

\subsection{Learning-based Quad Mesh Generation Method}
In recent years, learning-based quadrilateral remeshing~\cite{siddiqui2024meshgpt,liu2025quadgpt,dong2025neurcross,dong2025crossgen,liu2025neuframeq} has gained significant traction, primarily evolving along two distinct technical trajectories. The first category treats mesh faces or vertices as discrete tokens, utilizing autoregressive models for direct mesh synthesis. MeshGPT~\cite{siddiqui2024meshgpt} pioneered this paradigm by modeling triangular meshes as sequences of face tokens. QuadGPT~\cite{liu2025quadgpt} extended this approach to the quadrilateral domain, introducing the first end-to-end autoregressive model for generating native quad-dominant meshes. However, due to the inherent sequence-length bottlenecks of transformers, these methods struggle to produce high-resolution, high-precision meshes. Furthermore, they lack explicit control over topological edge flow, often resulting in fragmented or structurally inconsistent outputs.

The second trajectory integrates deep learning with traditional geometric optimization to predict coherent orientation fields. NeurCross~\cite{dong2025neurcross} adopts a self-supervised framework, transforming remeshing into a continuous energy minimization process by jointly optimizing neural implicit SDFs and cross-fields. CrossGen~\cite{dong2025crossgen} proposes a data-driven framework that models geometry and cross-fields within a shared latent space to achieve cross-category generalization. Furthermore, NeuFrameQ~\cite{liu2025neuframeq} utilizes frame fields as the core representation, employing regression networks and diffusion models to predict field orientation and magnitude, respectively.

Despite these advancements, existing methods lack explicit control over structural layouts, often leading to jagged boundaries, lost details, and poor editability. To address this, we propose simultaneously predicting structural lines and cross-fields. By learning semantic priors from large-scale data, our approach generates quad-meshes with high-quality topologies that meet industry standards.

\section{The TopGen-220K Dataset}
\begin{figure}[h]
    \centering
    \includegraphics[width=\textwidth]{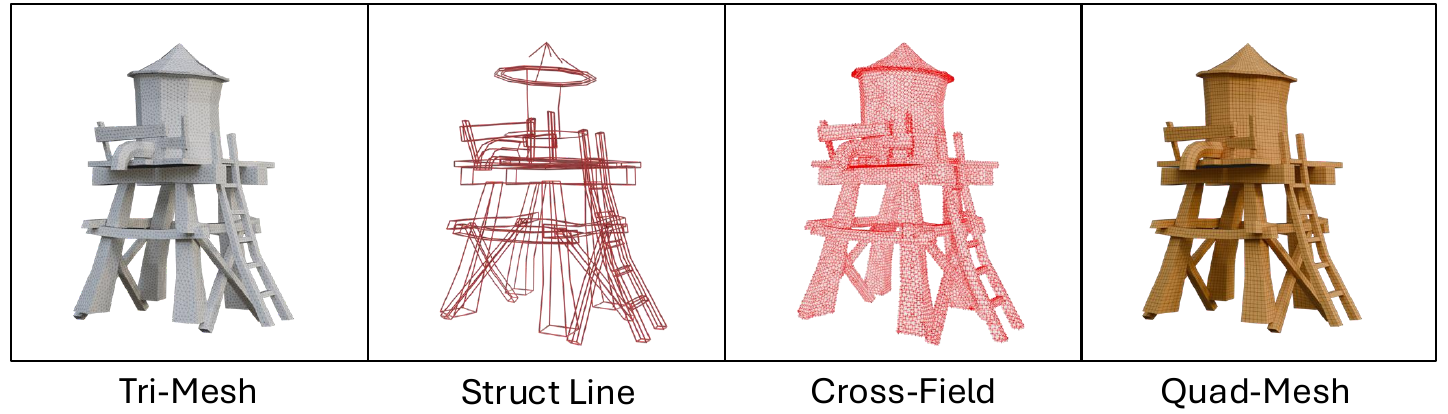}
    \caption{\textbf{A sample from TopGen-220K dataset}. From left to right: the input triangular mesh, the structural layout, the cross-field, and the quadrilateral mesh.}
    \label{fig:datasets}
\end{figure}

Constructing a structural layout is a pivotal step in professional artist-driven quadrilateral modeling. However, as these layouts are typically intermediate artifacts that are rarely preserved, such data remains critically scarce in existing research. To support our proposed joint-prediction framework for cross-fields and structure lines, we constructed TopGen-220K, a large-scale, diverse dataset comprising paired data of raw triangular meshes, structural layouts, cross-fields, and quadrilateral meshes (Figure~\ref{fig:datasets}).
We aggregated 1.3 million raw meshes from ShapeNetV2~\cite{chang2015shapenet}, 3D-FUTURE~\cite{fu20213d}, Objaverse~\cite{deitke2023objaverse}, Objaverse-XL~\cite{deitke2023objaversexl}, and proprietary licensed sources, utilizing the FlowRep~\cite{gori2017flowrep} to extract initial structure lines. For cross-field computation, we adopted the approach from ~\cite{dielen2021learning} for quadrilateral regions. For non-quadrilateral regions, we followed the methodology of QuadWild~\cite{pietroni2021reliable}, aligning the cross-field with the structure lines in adjacent areas and employing interpolation-diffusion for internal regions. Finally, using these structure lines and cross-fields as initial constraints, we optimized the final quadrilateral results via the QuadWild framework.
\begin{wrapfigure}{r} {0.45\textwidth}
\centering
\begin{center}
\includegraphics[width=\linewidth]{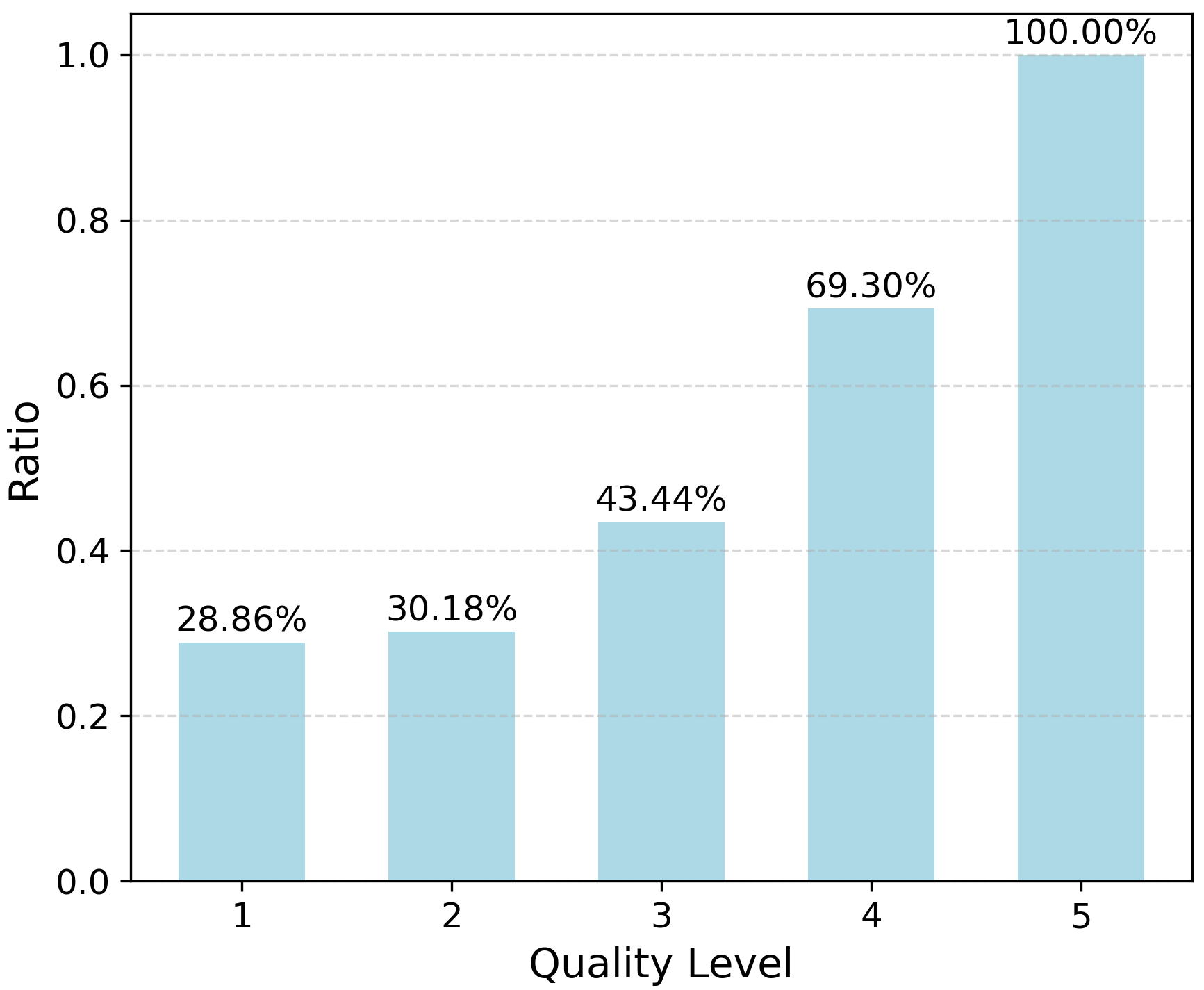}
\end{center}
\caption{Distribution of TopGen-220K.}
\label{fig:dataset_distribution}
\end{wrapfigure}
To ensure data quality, we recruited professional modelers to perform rigorous manual filtering based on the structural rationality and edge flow of the meshes. We excluded cases with cluttered, fragmented, or indiscernible structure lines, resulting in a curated set of 220,000 samples. As shown in Figure~\ref{fig:dataset_distribution}, these are categorized into five quality tiers:
\begin{itemize}
\item \textbf{Level 1}: Both structural lines and edge flow are of high quality.
\item \textbf{Level 2}: High-quality edge flow with fewer than 5 structural layout issues.
\item \textbf{Level 3}: Acceptable edge flow with fewer than 5 structural layout issues.
\item \textbf{Level 4}: High-quality edge flow with fewer than 5 structural layout issues.
\item \textbf{Level 5}: Acceptable edge flow with fewer than 10 structural layout issues.
\end{itemize}
Data from Levels 1 through 3 were selected for the training set, ultimately forming the TopGen-220K dataset.

\section{Method}
\label{sec:method}
\begin{figure}[h]
    \centering
    \includegraphics[width=\textwidth]{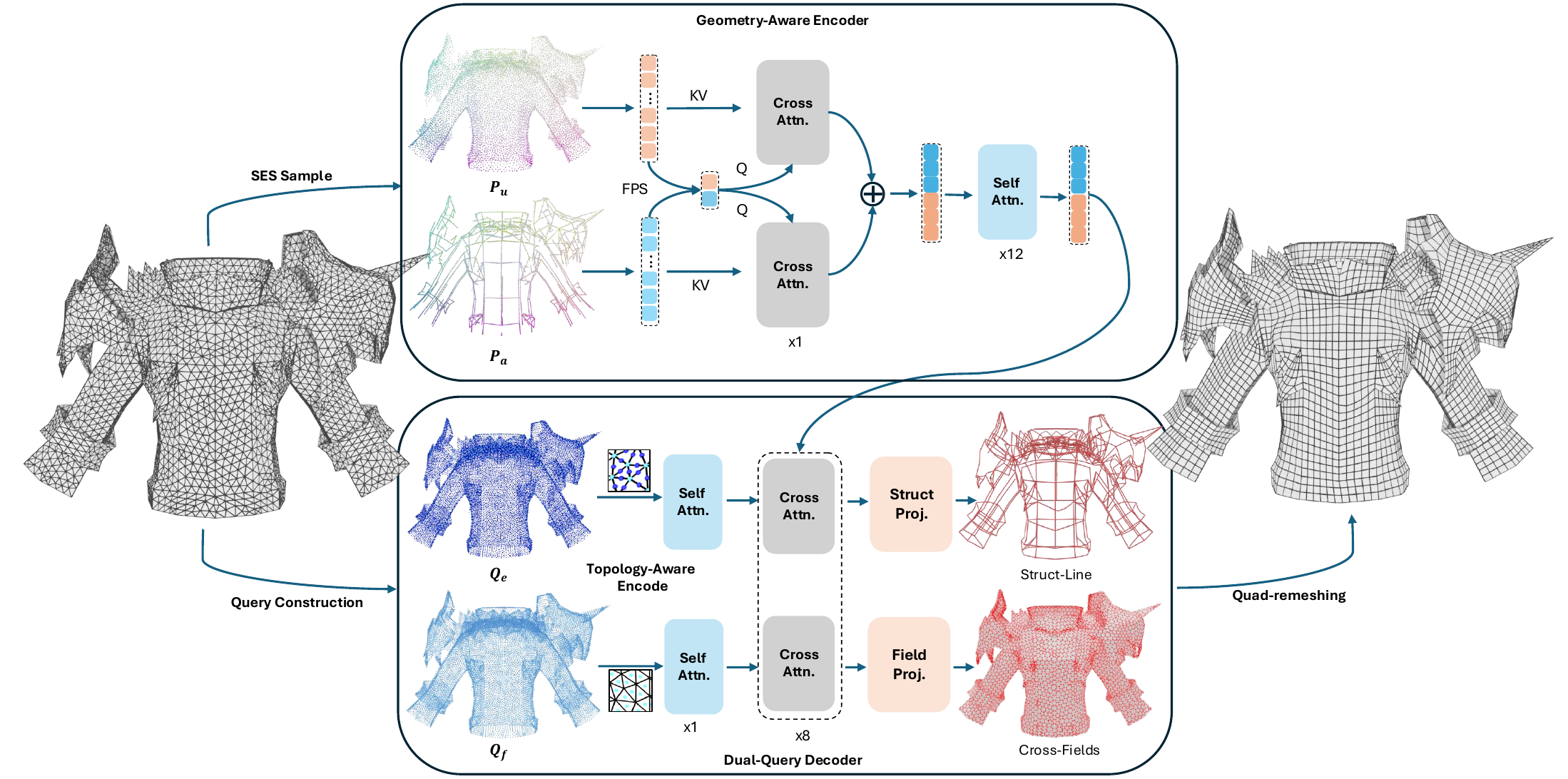}
    \caption{\textbf{Overview of the TopGen pipeline.} Our framework first samples point clouds from the input mesh via the SES strategy~\cite{chen2025dora}. A geometry-aware encoder then maps these points into a latent space. Subsequently, our proposed Dual-Query Decoder concurrently decodes the structural layouts and cross-fields in parallel. Finally, both predictions serve as joint guidance to facilitate high-quality quadrilateral remeshing.}
    \label{fig:pipeline}
\end{figure}
We propose TopGen, the first method that achieves quadrilateral mesh generation through the joint prediction of structural layouts and cross-fields. As illustrated in Figure~\ref{fig:pipeline}, our framework consists of three main components: a Geometry-Aware Encoder (Sec.~\ref{sec:encoder}) , a Dual-Query Decoder (Sec.~\ref{sec:decoder}) as well as a Structural-Lines and Cross-fields Driven Remeshing (Sec.~\ref{sec:remeshing}).
\subsection{Geometry-Aware Encoder}
\label{sec:encoder}
We adopt point clouds as the fundamental representation for our framework. As a flexible and expressive 3D represention, point clouds have become the predominant choice in modern deep-learning-based 3D vision tasks~\cite{chen2025dora,zhao2023michelangelo,zhang20233dshape2vecset}. Unlike mesh-based representations that are constrained by fixed connectivity, point clouds offer the versatility to handle diverse and potentially low-quality input topologies. Furthermore, this choice allows us to leverage powerful pre-trained point cloud encoders~\cite{chen2025dora,zhao2023michelangelo}. By utilizing backbones trained on massive 3D datasets, our framework inherits robust geometric encoding capabilities and strong shape priors, which are essential for high-fidelity reconstruction.

To preserve critical topological boundary information while extracting high-fidelity geometric features, we adopt a \emph{Sharp Edge Sampling} (SES) strategy and the \emph{Dual Cross-Attention} (DCA) architecture inspired by Dora-VAE~\cite{chen2025dora}. This paradigm enables simultaneous perception of global shape semantics and aggregation of fine-grained local geometric details.

\paragraph{\textbf{Sharp Edge Sampling.}}
Given an input mesh $\mathcal{M} = (\mathcal{V}, \mathcal{E}, \mathcal{F})$, the SES strategy identifies a set of salient edges $\Gamma$ by computing dihedral angles $\tau$ between adjacent faces. High-density adaptive sampling is then performed around these sharp features to generate a salient point set $\mathcal{P}_a$ consisting of $N_{desire}$ points, which effectively captures discriminative geometric characteristics of the shape. In parallel, a set of uniformly distributed surface samples $\mathcal{P}_u$ also comprising $N_{desire}$ points, is employed as a complementary representation. The union $\mathcal{P}_d = \mathcal{P}_a \cup \mathcal{P}_u$ forms a comprehensive and dense point cloud, ensuring thorough coverage of both fine-scale details and the overall geometric structure.
\paragraph{\textbf{Dual Cross-Attention Mechanism.}}
Within the DCA framework, Farthest Point Sampling (FPS) is first applied independently to $\mathcal{P}_a$ and $\mathcal{P}_u$ to generate a unified set of sparse anchor points $\mathcal{P}_s$:
\begin{equation}
\mathcal{P}_s = \operatorname{FPS}(\mathcal{P}_u, N_u) \;\cup\; \operatorname{FPS}(\mathcal{P}_a, N_a)
\end{equation}
where $\mathcal{N}_u$ and $\mathcal{N}_a$ is the number of downsampled point clouds from $\mathcal{P}_u$ and $\mathcal{P}_a$ respectively. These anchors serve as queries and are processed through parallel cross-attention layers with the salient point set $\mathcal{P}_s$ and the uniform point set $\mathcal{P}_u$, respectively, yielding decoupled feature embeddings $\mathcal{C}_u$ and $\mathcal{C}_a$. By fusing these specialized representations, the encoder obtains an intermediate feature $\mathcal{C}$ that encapsulates the multidimensional geometric properties of the input mesh:
\begin{equation}
\left\{
\begin{aligned}
\mathcal{C}_u &= \mathrm{CrossAttn}(\mathcal{P}_s, \mathcal{P}_u, \mathcal{P}_u) \\
\mathcal{C}_a &= \mathrm{CrossAttn}(\mathcal{P}_s, \mathcal{P}_a, \mathcal{P}_a)
\end{aligned}
\right.
\end{equation}
\begin{equation}
\mathcal{C} = \mathcal{C}_u + \mathcal{C}_a , \quad\mathcal{Z} = SelfAttn(\mathcal{C}) 
\end{equation}
Subsequently, $\mathcal{C}$ is further refined by a self-attention layer to model global feature interactions, producing the final latent feature $\mathcal{Z}$. This hierarchical feature extraction strategy provides rich prior information for subsequent decoding tasks.
\subsection{Dual-Query Decoder}
\label{sec:decoder}

The Dual-Query Decoder is the core component of TopGen, designed to jointly predict structural lines and cross fields in a unified and topology-aware manner. 

\begin{wrapfigure}{r} {0.5\textwidth}
\centering
\begin{center}
\includegraphics[width=\linewidth]{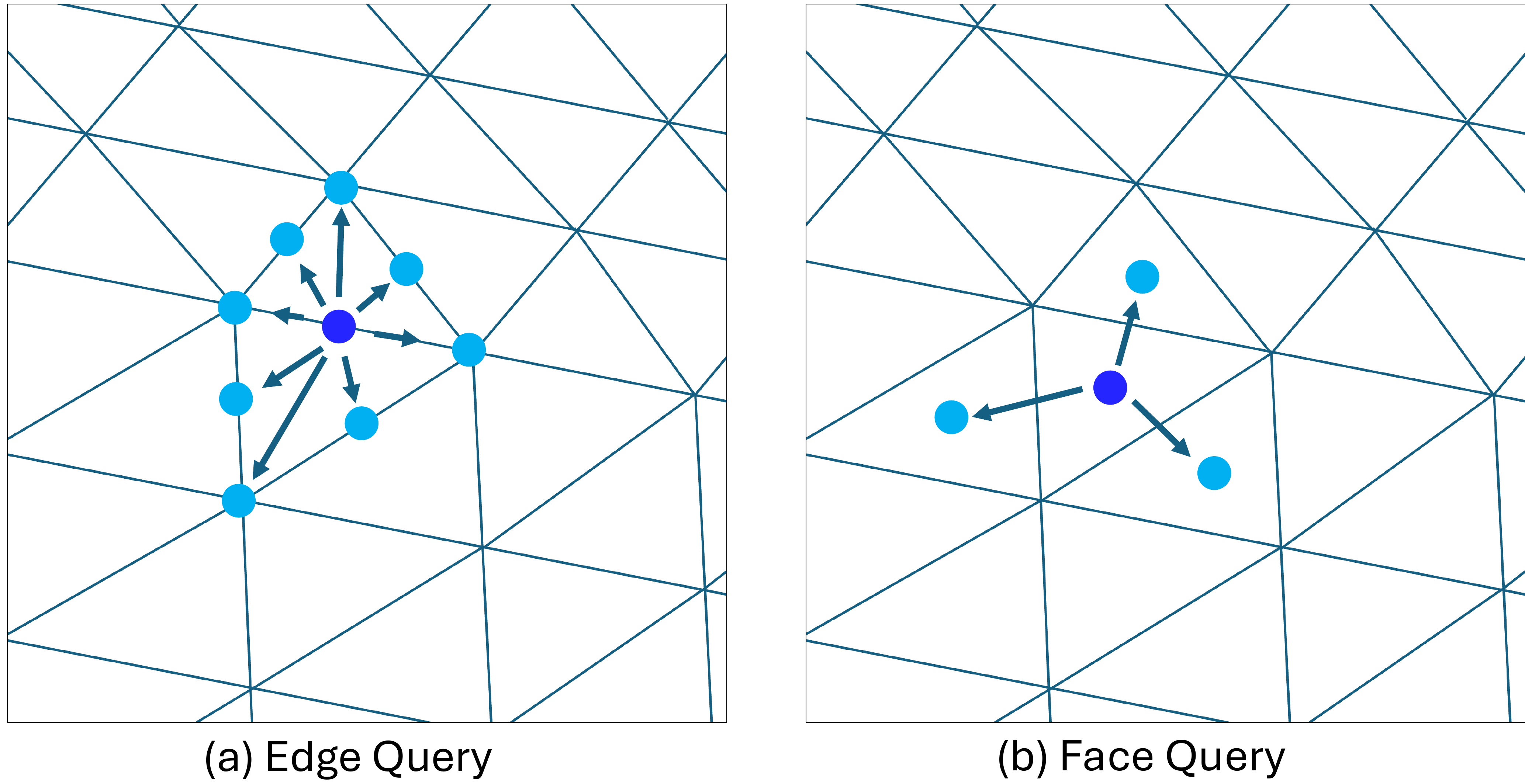}
\end{center}
\caption{Topology-aware neighborhoods.}
\label{fig:local_encode}
\end{wrapfigure}

\paragraph{\textbf{Mesh-Aligned Query Construction.}}
We construct two types of query points that are explicitly aligned with mesh primitives. Specifically, for each face $f \in \mathcal{F}$, we place a face query point at its barycenter, forming the set $\mathcal{Q}_f$. Likewise, for each edge $e \in \mathcal{E}$, we place an edge query point at its midpoint, forming the set $\mathcal{Q}_e$. This design establishes a one-to-one correspondence between queries and mesh elements, thereby fully preserving the original mesh topology throughout decoding.
\paragraph{\textbf{Topology-Aware Local Context Encoding.}}
Although each query point is associated with a specific mesh primitive, reliable prediction of structural layouts and cross fields requires rich local geometric context. As shown in Figure~\ref{fig:local_encode}, we therefore construct topology-aware neighborhoods for edge and face queries by explicitly exploiting mesh connectivity.
For an edge query point $q_e \in \mathcal{Q}_e$ corresponding to an edge $e \in \mathcal{E}$, we first collect all faces incident to $e$:
\begin{equation}
\mathcal{F}_e(e) = \{ f \in \mathcal{F} \mid e \subset f \}.
\end{equation}

The edge-topological neighborhood $\mathcal{N}_e$ is then defined as the union of all vertices and edge midpoints belonging to these incident faces:
\begin{equation}
\mathcal{N}_e =
\bigcup_{f \in \mathcal{F}_e(e)}
\left( \mathcal{V}(f) \cup \mathcal{E}_m(f) \right),
\end{equation}
where $\mathcal{V}(f)$ and $\mathcal{E}_m(f)$ denote the vertex set and the set of edge midpoints of face $f$, respectively. This neighborhood encodes both local edge connectivity and surrounding surface geometry, which are crucial for identifying structural lines.

Similarly, for a face query point $q_f \in \mathcal{Q}_f$ corresponding to a face $f \in \mathcal{F}$, we define its adjacent face set as
\begin{equation}
\mathcal{F}_f(f) = \{ f' \in \mathcal{F} \mid f' \text{ shares an edge with } f \}.
\end{equation}
The face-topological neighborhood $\mathcal{N}_f$ is constructed by collecting the barycenters of all adjacent faces:
\begin{equation}
\mathcal{N}_f = \{ q_{f'} \mid f' \in \mathcal{F}_f(f) \},
\end{equation}
providing a coherent local surface context for cross-field estimation.

Each element in $\mathcal{N}_e$ and $\mathcal{N}_f$ is represented by its 3D coordinates and corresponding normal information. Edge and face queries aggregate their local topological context via attention-based feature aggregation:
\begin{equation}
\mathcal{X}_e = \mathrm{SelfAttn}(q_e, \mathcal{N}_e), \quad
\mathcal{X}_f = \mathrm{SelfAttn}(q_f, \mathcal{N}_f),
\end{equation}
yielding locally enhanced query features for subsequent decoding.

\paragraph{\textbf{Structural Line Decoding.}}
The enhanced edge query features $\mathcal{X}_e$ are decoded by attending to the encoder latent representation $\mathcal{Z}$ through cross-attention. Structural line prediction is formulated as a binary classification task, where each edge query outputs the probability that the corresponding edge belongs to the structural layout. Since sharp edges often correspond to structural features, we explicitly incorporate this prior by adding a bias term $W_{\text{prior}}$ to edge queries associated with salient edges $\Gamma$:
\begin{equation}
\mathcal{S} = \mathrm{CrossAttn}(\mathcal{X}_e, \mathcal{Z}), \quad
\mathcal{S}_{pred} = \mathcal{S} + W_{\text{prior}}.
\end{equation}
The structural line branch is supervised using a binary cross-entropy loss:
\begin{equation}
\mathcal{L}_{\text{struct}} = \mathrm{BCE}(\mathcal{S}_{pred}, \mathcal{S}_{gt}).
\end{equation}

\paragraph{\textbf{Cross Field Decoding.}}
In parallel, the enhanced face query features $\mathcal{X}_f$ interact with the encoder latent features $\mathcal{Z}$ via a cross-attention mechanism to regress the cross-field. We employ a 2-rosy cross-field, where the field at each face is represented by a vector $(u, v)$. To resolve angular symmetry and facilitate network convergence, we follow the representation used in ~\cite{liu2025neuframeq} by mapping the cross-field into a polyvector space: $c_0 = -(u^2 + v^2)$ and $c_1 = u^2v^2$. Each face query directly predicts these coefficients defined on the corresponding face:
\begin{equation}
R_{pred} = (c_0, c_1) = \mathrm{CrossAttn}(\mathcal{X}_f, \mathcal{Z})
\end{equation}
Furthermore, inspired by NeurCross~\cite{dong2025neurcross}, we assign spatially varying weights to the face queries based on their proximity to salient edges $\Gamma$. Queries closer to $\Gamma$ are assigned higher weights to enforce sharp feature alignment. Specifically, the weighting factor $D_p$ is defined as:
\begin{equation}
D_{p} = 1 - e^{-\rho_{feature} \cdot d(p, \Gamma)}
\end{equation}

where $d(p, \Gamma)$ denotes the geodesic distance between the face query point $p$ and the nearest salient edge $\Gamma$, with $\rho_{feature}$ set to 10 by default. The final cross-field objective is supervised using a weighted Mean Squared Error loss:
\begin{equation}
\mathcal{L}_{crossfield} = D_p \cdot \|R_{pred} - R_{gt}\|^2_2
\end{equation}

The final training objective is a weighted sum of the two task-specific losses:
\begin{equation}
\mathcal{L}_{\text{total}} =
w_{\text{struct}}\mathcal{L}_{\text{struct}} +
w_{\text{crossfield}}\mathcal{L}_{\text{crossfield}}
\label{eq:total_loss}
\end{equation}

\subsection{Structural-Lines and Cross-fields Driven Quad-Remeshing}
\label{sec:remeshing}
Upon predicting structural lines and cross-fields, we integrate them into a robust remeshing pipeline inspired by QuadWild~\cite{pietroni2021reliable}. By decoupling learned semantic guidance from the classical optimization backend, TopGen leverages the stability of traditional methods while achieving superior topological rationality.

\paragraph{\textbf{Constraint Integration.}}
We threshold the predicted probabilities to extract structural edges, which naturally form polylines respecting the original mesh topology. These are enforced as hard feature constraints to prevent edge flips or feature drift. Simultaneously, the predicted 2-RoSy cross-fields serve as soft orientation constraints, defining a globally consistent directional field to guide field-aligned parameterization.

\paragraph{\textbf{Quad Extraction.}}
Quad elements are generated by tracing integer iso-lines in the parameter domain. By enforcing structural lines as rigid boundaries and cross-fields as orientation guides, the resulting mesh faithfully captures sharp geometric features and semantic structures while maintaining high regularity and low distortion.

\section{Experiments}
\subsection{Implementation Details}

\paragraph{\textbf{Sampling}} we sample $N_{desire} = 16,384$ points with a dihedral angle threshold of $\tau = 30^{\circ}$. And we apply FPS to obtain $N_u = N_a = 2,048$ points, effectively balancing uniform surface coverage with feature-aligned samples. The number of query points is set to $N_e = N_f = 61,440$. To enhance training stability for structural line prediction, we adopt a $1:2$ ratio for positive and negative samples during structural query point sampling.

\paragraph{\textbf{Network Architecture.}} The Geometry-Aware Encoder employs a latent width of $768$ with $12$ attention heads. It comprises one parallel cross-attention layer for feature interaction, followed by $12$ self-attention layers for global context extraction. The Dual-Query Decoder begins with a single-layer Topology-Aware Local Context Encoding, followed by $8$ cross-attention layers to jointly decode structural layouts and cross-fields.

\paragraph{\textbf{Training Configuration.}} TopGen was trained on $16$ NVIDIA A100 GPUs for $30,000$ steps with a batch size of $8$ and a learning rate of $5 \times 10^{-5}$. We initialized the encoder with pre-trained Dora-VAE~\cite{chen2025dora} weights. Optimization follows the objective in Eq.~\ref{eq:total_loss} with hyper-parameters $w_{\text{crossfield}} = 1.0$, $w_{\text{struct}} = 0.1$, and a structural prior bias $W_{\text{prior}} = 0.2$.

\subsection{Comparison With Quadrilateral Method}
\label{sec:method}
We conduct a comprehensive comparative analysis against state-of-the-art open-source quadrilateral remeshing techniques, including Instant-Meshes~\cite{jakob2015instant}, QuadriFlow~\cite{huang2018quadriflow}, and QuadWild~\cite{pietroni2021reliable}. Evaluations are performed on the QuadWild-300 dataset and a collection of AI-generated models.
\begin{figure}[h]
    \centering
    \includegraphics[width=\textwidth]{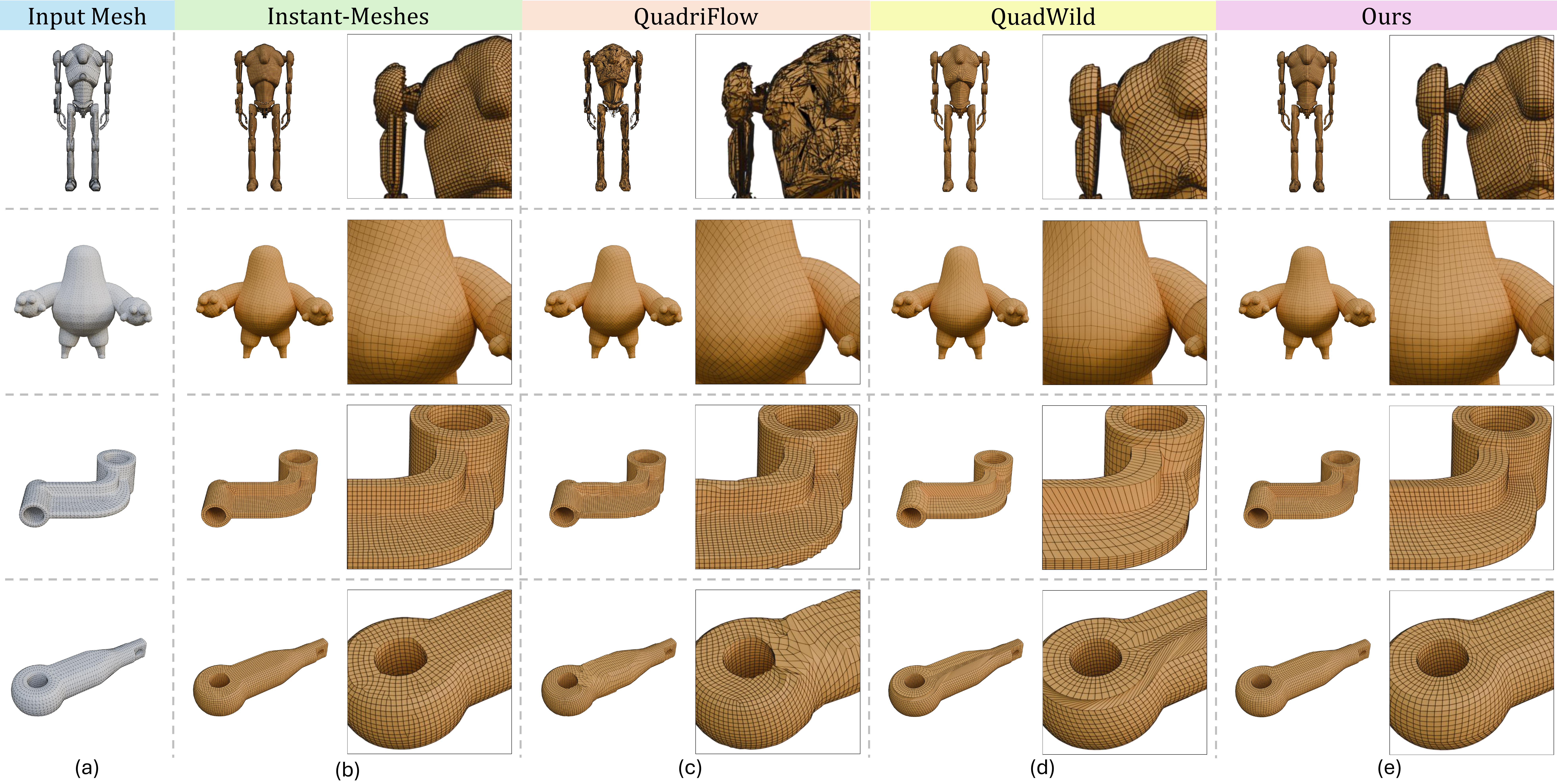}
    \caption{\textbf{Qualitative comparison with quadrilateral remeshing methods.} (a) Input meshes: AI-generated (Robot, Monster) and QuadWild-300 samples (Mech09, Rod). (e) Results from Instant-Meshes, QuadriFlow, ,QuadWild and Our TopGen.}
    \label{fig:Qualitative_result}
\end{figure}
The qualitative comparisons in Figure~\ref{fig:Qualitative_result} reveal distinct limitations in existing methods. While Instant-Meshes minimizes local distortion, it frequently introduces an excessive density of singularities, leading to chaotic global edge flows. QuadriFlow exhibits significant instability, often suffering from geometric corruption and detail loss when encountering non-manifold topologies or high-genus models. Similarly, QuadWild struggles to maintain layout consistency across organic surfaces and complex topologies. In contrast, our method consistently synthesizes high-quality topologies across a vast spectrum of inputs—including smooth organic forms, sharp mechanical surfaces, and challenging non-manifold geometries—yielding results that closely align with professional manual modeling standards. More results are provided in Figure~\ref{fig:teaser}

\begin{table}[h]
\caption{\textbf{Quantitative comparison with quadrilateral remeshing method.} Best results are bolded and underlined; second-best are bolded. Our method consistently achieves superior geometric fidelity and topological quality without structural corruption, outperforming state-of-the-art methods in robustness.}
\label{tab:quantitative_results}
\centering
\resizebox{\textwidth}{!}{ % 缩放到整页宽度
\small
\renewcommand{\arraystretch}{1.2}
\begin{tabular}{l|cccccc|cccccc}
\toprule
\multicolumn{1}{c}{} & 
\multicolumn{6}{|c}{Robot} & 
\multicolumn{6}{|c}{Monster} \\
\multicolumn{1}{c|}{Method} & 
\hspace{0.4cm} V \hspace{0.4cm} & \hspace{0.4cm} F \hspace{0.4cm} & \hspace{0.4cm} I$\downarrow$ \hspace{0.4cm} & \hspace{0.1cm} CD$\times \left( 10^{-5} \right )\downarrow$ \hspace{0.1cm} & \hspace{0.4cm} SJ$\downarrow$ \hspace{0.4cm} & \hspace{0.2cm} Corruption \hspace{0.2cm} &
\hspace{0.4cm} V \hspace{0.4cm} & \hspace{0.4cm} F \hspace{0.4cm} & \hspace{0.4cm} I$\downarrow$ \hspace{0.4cm} & \hspace{0.1cm} CD$\times \left( 10^{-5} \right )\downarrow$ \hspace{0.1cm} & \hspace{0.4cm} SJ$\downarrow$ \hspace{0.4cm} & \hspace{0.2cm} Corruption \hspace{0.2cm}\\
\hline
Instant-Meshes  & 36057 & 30717 & 8946 & 4.278 & 0.395 & \checkmark & 4745 & 4698 & 367 & \textbf{48.463} & 0.475 & \texttimes \\
QuadriFlow      & 37286 & 32505 & 8306 & 4.102 & \underline{\textbf{0.117}} & \checkmark & 4558 & 4495 & \underline{\textbf{189}} & 64.33 & 0.460 & \texttimes \\
Quadwild        & 19482 & 17144 & \underline{\textbf{5683}} & \textbf{4.044} & 0.354 & \checkmark & 5368 & 5275 & 234 & 52.281 & \underline{\textbf{0.429}} & \texttimes \\
Ours            & 35890 & 32391 & \textbf{8163} & \underline{\textbf{3.300}} & \textbf{0.352} & \texttimes & 5293 & 5204 & \textbf{212} & \underline{\textbf{44.404}} & \textbf{0.454} & \texttimes \\
\hline
\multicolumn{1}{c}{} & 
\multicolumn{6}{|c}{Mech09} & 
\multicolumn{6}{|c}{Rod} \\
\multicolumn{1}{c|}{Method} & 
V & F & I$\downarrow$ & CD$\times \left( 10^{-5} \right )\downarrow$ & SJ$\downarrow$ & Corruption &
V & F & I$\downarrow$ & CD$\times \left( 10^{-5} \right )\downarrow$ & SJ$\downarrow$ & Corruption \\
\hline
Instant-Meshes  & 8406 & 8433 & 218 & \textbf{6.762} & \textbf{0.336} & \texttimes & 4428 & 4443 & 133 & \textbf{8.924} & 0.497 & \texttimes \\
QuadriFlow      & 8788 & 8790 & 66 & 6.778 & \underline{\textbf{0.334}} & \checkmark & 4296 & 4298 & \textbf{42} & 9.094 & \underline{\textbf{0.474}} & \texttimes \\
Quadwild        & 3593 & 3595 & \textbf{22} & 11.578 & 0.350 & \texttimes & 4143 & 4145 & \underline{\textbf{24}} & 9.397 & 0.493 & \texttimes \\
Ours            & 8274 & 8276 & \underline{\textbf{16}} & \underline{\textbf{6.721}} & 0.372 & \texttimes & 4443 & 4445 & \underline{\textbf{24}} & \underline{\textbf{8.372}} & \textbf{0.487} & \texttimes \\
\bottomrule
\end{tabular}
}
\end{table}

Furthermore, we provide a quantitative evaluation of the models from Figure~\ref{fig:Qualitative_result} in Table~\ref{tab:quantitative_results}. We report performance across six key metrics: vertex count (V), face count (F), number of singularities (I), Chamfer Distance (CD), average Scaled Jacobian (SJ)~\cite{liu2025neuframeq}, and geometric corruption (Corruption). Specifically, we define Corruption as the presence of visible structural fractures or broken faces, which are identified through meticulous visual inspection. Notably, our method achieves the lowest Chamfer Distance across all test cases, underscoring its superior ability to preserve geometric fidelity. Compared to competing methods, our approach maintains the fewest singularities and a low average Scaled Jacobian at comparable face counts, ensuring high-quality topology and geometry without any corruption.

\subsection{Comparison With FlowRep}
\begin{figure}[h]
    \centering
    \includegraphics[width=\textwidth]{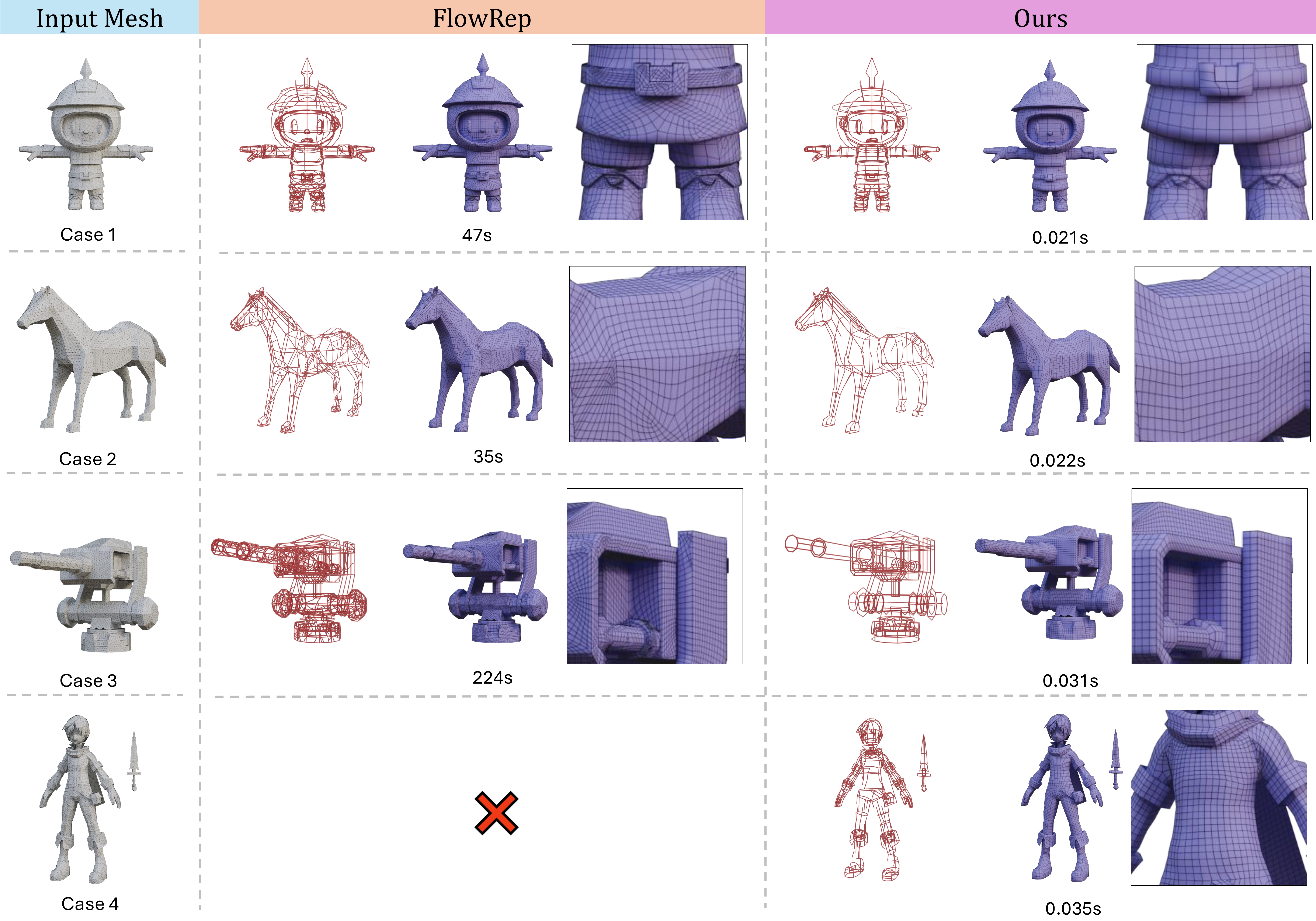}
    \caption{\textbf{Comparison with FlowRep.} From left: Input, FlowRep, and TopGen. We visualize predicted structure lines and final quad-meshes with timing reported below.}
    \label{fig:struct_result}
\end{figure}
We selected models of varying geometric complexity to evaluate our predicted structure lines against the baseline FlowRep. To ensure a rigorous and controlled comparison, we utilized our model-predicted cross-field as the common guidance for both methods, and applied the identical remeshing pipeline (detailed in Sec.~\ref{sec:remeshing}) to generate the final quadrilateral meshes. This experimental setup allows for a direct assessment of how the quality and topology of the structure lines themselves influence the final remeshing outcome.

As shown in Figure~\ref{fig:struct_result}, FlowRep exhibits instability on non-manifold geometries (Case 4) and often generates redundant lines that clutter the final edge flow (Cases 1, 2). Furthermore, its optimization-based nature leads to significant latency on high-resolution models (Case 3). In contrast, our model leverages learned geometric priors to produce concise, semantically meaningful structure lines across diverse topologies. Most notably, our encoder-decoder architecture reduces processing time from hundreds of seconds to a average of 0.0x seconds.

\subsection{Ablation Study}
\begin{figure}[h]
    \centering
    \includegraphics[width=\textwidth]{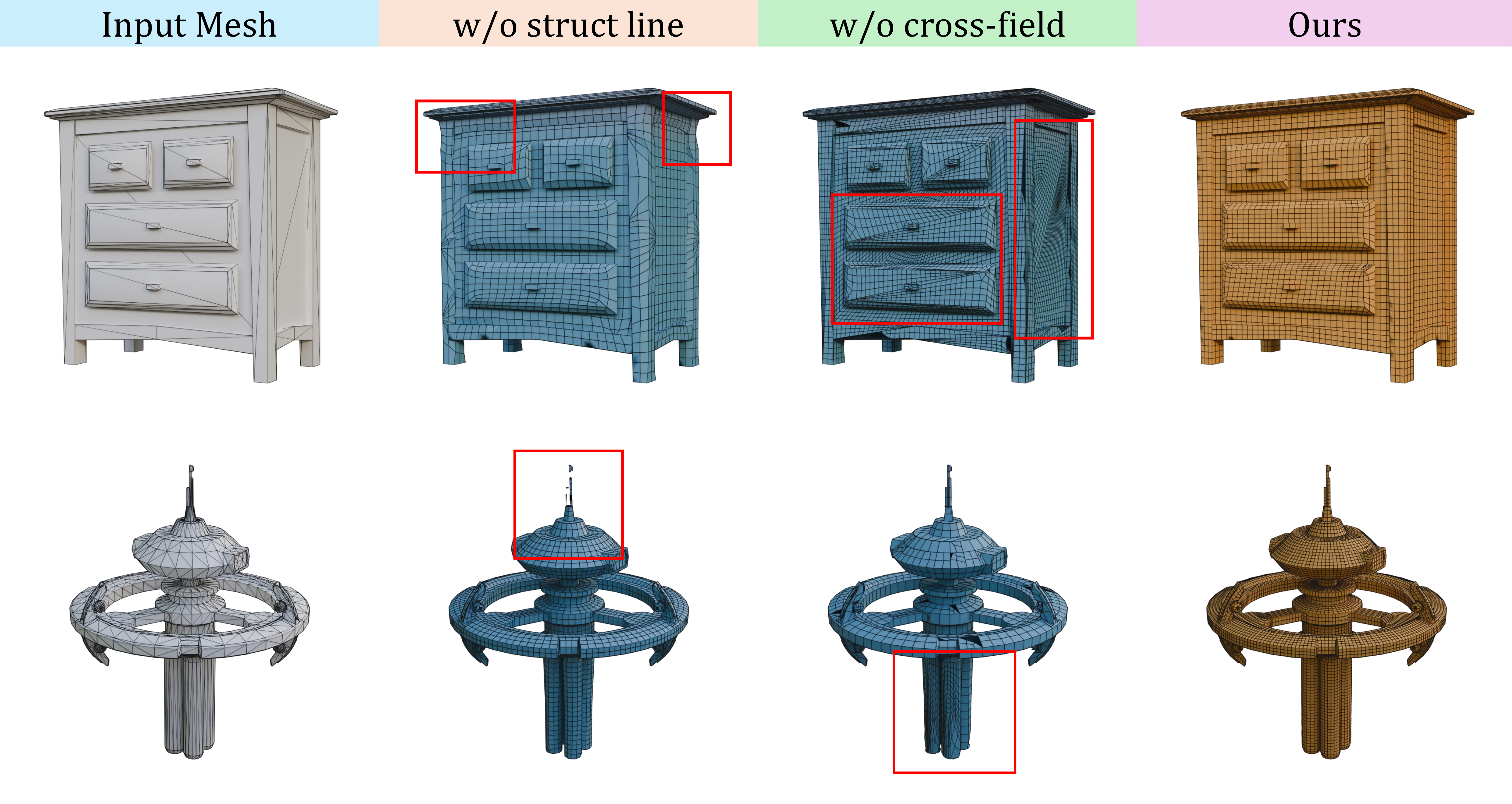}
    \caption{\textbf{Ablation results.} Red boxes highlight artifacts from omitting structure lines (jagged boundaries and detail loss) or cross-fields (irregular edge flow and topological corruption).}
    \label{fig:ablation}
\end{figure}
We evaluate the necessity of our joint-prediction framework by comparing it against two variants: (i) guidance solely by structure lines and (ii) guidance exclusively by cross-fields.
As shown in Figure~\ref{fig:ablation}, relying only on cross-fields lacks global layout constraints, resulting in jagged boundaries and geometric detail loss. Conversely, using only structure lines maintains the external framework but leads to disorganized internal edge flows due to the absence of orientation guidance. These results confirm that joint prediction is essential for simultaneously ensuring boundary integrity and internal topological regularity.

\section{Conclusion}
In this paper, we present TopGen, the first quadrilateral mesh generation framework that jointly predicts structural layouts and cross-fields. By leveraging the synergistic guidance of structural lines and cross-fields, TopGen synthesizes quad meshes that not only preserve intricate geometric details but also exhibit high-quality, artist-standard topological edge flows. Furthermore, we introduce TopGen-220K, a large-scale, high-quality dataset featuring curated structural lines and cross-field annotations, effectively filling the long-standing void of structural feature data in the quad-remeshing community. Extensive experiments demonstrate that TopGen significantly outperforms state-of-the-art methods, offering substantial research value and practical utility for downstream graphics pipelines.

\appendix

\section{Appendix}

\subsection{Comparison With NeurCross}
\begin{figure}[h]
    \centering
    \includegraphics[width=\textwidth]{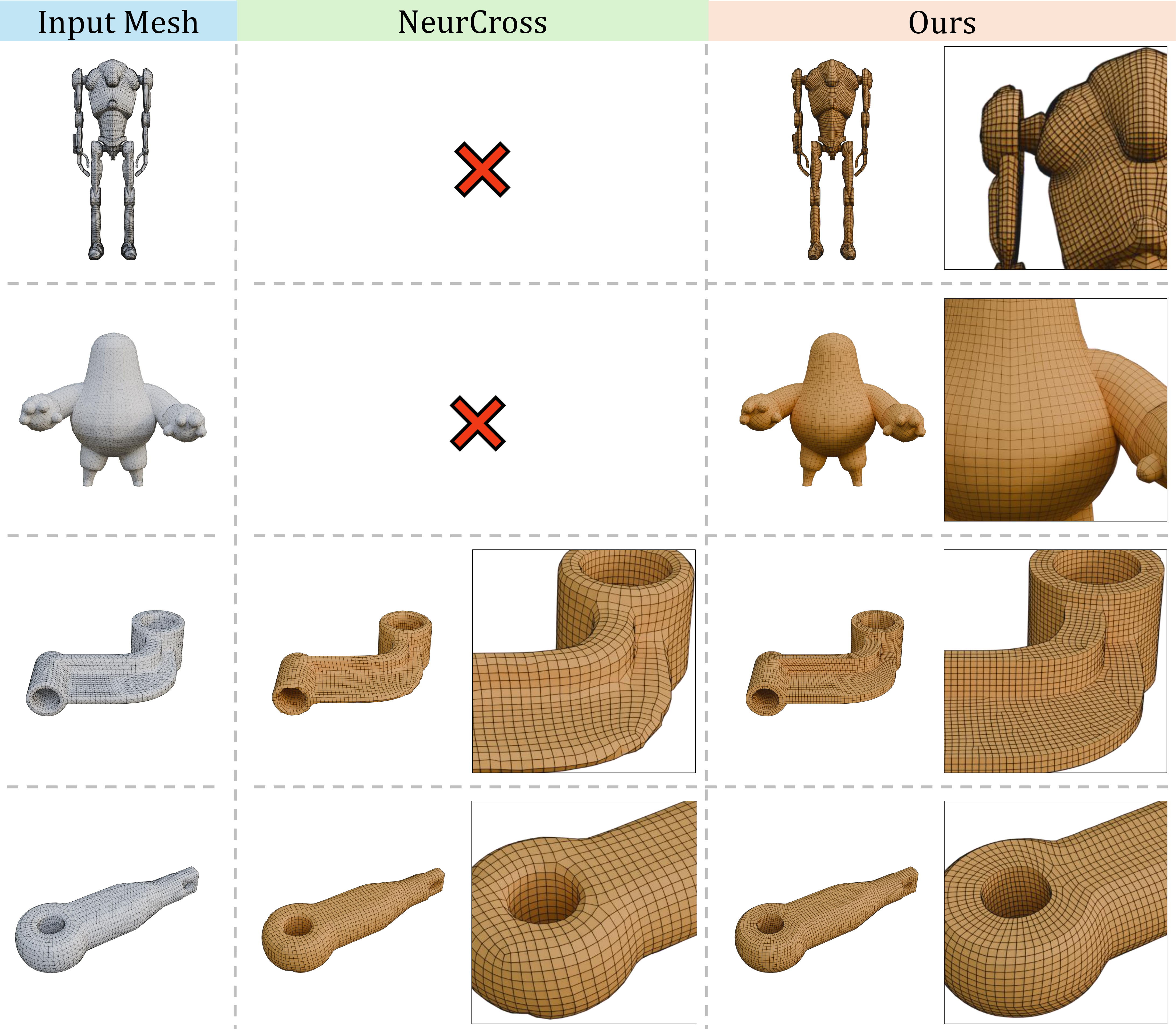}
    \caption{\textbf{Qualitative comparison with NeurCross.} (Top to bottom) AI-generated models (Robot, Monster) and QuadWild-300 samples (Mech09, Rod). While NeurCross exhibits numerical instability on AI-generated inputs, our method consistently achieves superior results, ensuring both precise external geometric profiles and smooth internal edge flows across all sources.}
    \label{fig:neurcross}
\end{figure}

We conduct an additional experiment to evaluate our method against NeurCross~\cite{dong2025neurcross}. For a direct and intuitive comparison, we present qualitative (Figure~\ref{fig:neurcross}) and quantitative (Table~\ref{tab:neurcross}) results using the same models as in the main manuscript. We utilize the official open-source implementation of NeurCross with default parameters. Since NeurCross exclusively predicts cross-fields, we apply the identical quad-remeshing pipeline (Sec.~\ref{sec:remeshing}) to its outputs for a fair assessment.

Notably, NeurCross encountered training divergence when processing AI-generated models (Robot and Monster), where the predicted cross-fields resulted in NaN values, precluding the generation of final quad-meshes. On the QuadWild-300 samples (Mech09 and Rod), while NeurCross produces reasonable internal edge flows through its cross-field optimization, the absence of explicit structural constraints leads to jagged boundaries and a loss of geometric detail. Furthermore, as an optimization-based approach requiring per-model fitting, NeurCross suffers from prohibitively long processing times, which is often impractical for professional production pipelines.

\begin{table}[h]
\caption{\textbf{Quantitative comparison with NeurCross.} Compared to NeurCross, our method achieves superior geometric fidelity, fewer singularities and lower average Scaled Jacobian values. Notably, while NeurCross requires extensive optimization time (>1,000s), our approach performs near-instantaneous inference (<1s).}
\label{tab:neurcross}
\centering
\resizebox{\textwidth}{!}{
\small
\renewcommand{\arraystretch}{1.2}
\begin{tabular}{l|ccccccc}
\toprule
\multicolumn{8}{c}{Mech09} \\
\hline
Method & \hspace{0.4cm} V \hspace{0.4cm} & \hspace{0.4cm} F \hspace{0.4cm} & \hspace{0.4cm} I$\downarrow$ \hspace{0.4cm} & \hspace{0.4cm} CD$\times (10^{-5})\downarrow$ \hspace{0.4cm} & \hspace{0.4cm} SJ$\downarrow$ \hspace{0.4cm} & \hspace{0.4cm} Corruption \hspace{0.4cm} & \hspace{0.4cm} Runtime(s)$\downarrow$ \hspace{0.4cm} \\
\hline
NeurCross & 3932 & 3934 & 30 & 8.892 & \textbf{0.365} & \texttimes & 1818\\
Ours      & 8274 & 8276 & \textbf{16} & \textbf{6.721} & 0.372 & \texttimes & \textbf{0.028}\\
\hline
\multicolumn{8}{c}{Rod} \\
\hline
Method & V & F & I$\downarrow$ & CD$\times (10^{-5})\downarrow$ & SJ$\downarrow$ & Corruption & Runtime(s)$\downarrow$\\
\hline
NeurCross & 2291 & 2293 & 34 & 21.93 & 0.490 & \texttimes & 1273\\
Ours      & 4443 & 4445 & \textbf{24} & \textbf{8.372} & \textbf{0.487} & \texttimes & \textbf{0.027}\\
\bottomrule
\end{tabular}
}
\end{table}

\subsection{Future Work}
While our method demonstrates robustness in predicting structural lines and cross-fields across diverse mesh sources, its performance on high-resolution models (e.g., those with millions of faces) still leaves room for improvement in terms of fine-grained precision. We believe these limitations can be addressed by incorporating a larger volume of authentic, human-authored modeling data into our training set. Furthermore, we have observed that the TopGen architecture holds significant potential for other layout-prediction tasks, such as UV unwrapping and part segmentation. We intend to explore these promising directions and further refine the scalability of our approach in future research.

\bibliographystyle{abbrv}
\bibliography{main}

@article{jakob2015instant,
  title={Instant field-aligned meshes.},
  author={Jakob, Wenzel and Tarini, Marco and Panozzo, Daniele and Sorkine-Hornung, Olga and others},
  journal={ACM Trans. Graph.},
  volume={34},
  number={6},
  pages={189--1},
  year={2015}
}

@inproceedings{huang2018quadriflow,
  title={Quadriflow: A scalable and robust method for quadrangulation},
  author={Huang, Jingwei and Zhou, Yichao and Niessner, Matthias and Shewchuk, Jonathan Richard and Guibas, Leonidas J},
  booktitle={Computer Graphics Forum},
  volume={37},
  number={5},
  pages={147--160},
  year={2018},
  organization={Wiley Online Library}
}

@article{bommes2009mixed,
  title={Mixed-integer quadrangulation},
  author={Bommes, David and Zimmer, Henrik and Kobbelt, Leif},
  journal={ACM transactions on graphics (TOG)},
  volume={28},
  number={3},
  pages={1--10},
  year={2009},
  publisher={ACM New York, NY, USA}
}

@article{pietroni2021reliable,
  title={Reliable feature-line driven quad-remeshing},
  author={Pietroni, Nico and Nuvoli, Stefano and Alderighi, Thomas and Cignoni, Paolo and Tarini, Marco and others},
  journal={ACM Transactions on Graphics},
  volume={40},
  number={4},
  pages={1--17},
  year={2021},
  publisher={Association for Computing Machinery}
}

@article{liang2025field,
  title={Field Smoothness-Controlled Partition for Quadrangulation},
  author={Liang, Zhongxuan and Du, Wei and Fu, Xiao-Ming},
  journal={ACM Transactions on Graphics (TOG)},
  volume={44},
  number={4},
  pages={1--15},
  year={2025},
  publisher={ACM New York, NY, USA}
}

@article{panozzo2014frame,
  title={Frame fields: Anisotropic and non-orthogonal cross fields additional material},
  author={Panozzo, Daniele and Puppo, Enrico and Tarini, Marco and Sorkine-Hornung, Olga},
  journal={Proceedings of the ACM TRANSACTIONS ON GRAPHICS (PROCEEDINGS OF ACM SIGGRAPH},
  volume={3},
  year={2014}
}

@article{gori2017flowrep,
  title={Flowrep: Descriptive curve networks for free-form design shapes},
  author={Gori, Giorgio and Sheffer, Alla and Vining, Nicholas and Rosales, Enrique and Carr, Nathan and Ju, Tao},
  journal={ACM Transactions on Graphics (TOG)},
  volume={36},
  number={4},
  pages={1--14},
  year={2017},
  publisher={ACM New York, NY, USA}
}

@inproceedings{siddiqui2024meshgpt,
  title={Meshgpt: Generating triangle meshes with decoder-only transformers},
  author={Siddiqui, Yawar and Alliegro, Antonio and Artemov, Alexey and Tommasi, Tatiana and Sirigatti, Daniele and Rosov, Vladislav and Dai, Angela and Nie{\ss}ner, Matthias},
  booktitle={Proceedings of the IEEE/CVF conference on computer vision and pattern recognition},
  pages={19615--19625},
  year={2024}
}

@article{liu2025quadgpt,
  title={QuadGPT: Native Quadrilateral Mesh Generation with Autoregressive Models},
  author={Liu, Jian and Wang, Chunshi and Guo, Song and Weng, Haohan and Zhou, Zhen and Li, Zhiqi and Yu, Jiaao and Zhu, Yiling and Xu, Jing and Lei, Biwen and others},
  journal={arXiv preprint arXiv:2509.21420},
  year={2025}
}

@article{dong2025neurcross,
  title={NeurCross: A Neural Approach to Computing Cross Fields for Quad Mesh Generation},
  author={Dong, Qiujie and Wen, Huibiao and Xu, Rui and Chen, Shuangmin and Zhou, Jiaran and Xin, Shiqing and Tu, Changhe and Komura, Taku and Wang, Wenping},
  journal={ACM Transactions on Graphics (TOG)},
  volume={44},
  number={4},
  pages={1--17},
  year={2025},
  publisher={ACM New York, NY, USA}
}

@article{dong2025crossgen,
  title={CrossGen: Learning and Generating Cross Fields for Quad Meshing},
  author={Dong, Qiujie and Wang, Jiepeng and Xu, Rui and Lin, Cheng and Liu, Yuan and Xin, Shiqing and Zhong, Zichun and Li, Xin and Tu, Changhe and Komura, Taku and others},
  journal={ACM Transactions on Graphics (TOG)},
  volume={44},
  number={6},
  pages={1--15},
  year={2025},
  publisher={ACM New York, NY, USA}
}

@inproceedings{liu2025neuframeq,
  title={NeuFrameQ: Neural Frame Fields for Scalable and Generalizable Anisotropic Quadrangulation},
  author={Liu, Ying-Tian and Li, Jiajun and Liu, Yu-Tao and Yu, Xin and Guo, Yuan-Chen and Cao, Yan-Pei and Liang, Ding and Shamir, Ariel and Zhang, Song-Hai},
  booktitle={Proceedings of the IEEE/CVF International Conference on Computer Vision},
  pages={28000--28009},
  year={2025}
}

@article{zhao2025hunyuan3d,
  title={Hunyuan3d 2.0: Scaling diffusion models for high resolution textured 3d assets generation},
  author={Zhao, Zibo and Lai, Zeqiang and Lin, Qingxiang and Zhao, Yunfei and Liu, Haolin and Yang, Shuhui and Feng, Yifei and Yang, Mingxin and Zhang, Sheng and Yang, Xianghui and others},
  journal={arXiv preprint arXiv:2501.12202},
  year={2025}
}

@article{xiang2025native,
  title={Native and compact structured latents for 3d generation},
  author={Xiang, Jianfeng and Chen, Xiaoxue and Xu, Sicheng and Wang, Ruicheng and Lv, Zelong and Deng, Yu and Zhu, Hongyuan and Dong, Yue and Zhao, Hao and Yuan, Nicholas Jing and others},
  journal={arXiv preprint arXiv:2512.14692},
  year={2025}
}

@article{chang2015shapenet,
  title={Shapenet: An information-rich 3d model repository},
  author={Chang, Angel X and Funkhouser, Thomas and Guibas, Leonidas and Hanrahan, Pat and Huang, Qixing and Li, Zimo and Savarese, Silvio and Savva, Manolis and Song, Shuran and Su, Hao and others},
  journal={arXiv preprint arXiv:1512.03012},
  year={2015}
}

@article{fu20213d,
  title={3d-future: 3d furniture shape with texture},
  author={Fu, Huan and Jia, Rongfei and Gao, Lin and Gong, Mingming and Zhao, Binqiang and Maybank, Steve and Tao, Dacheng},
  journal={International Journal of Computer Vision},
  volume={129},
  number={12},
  pages={3313--3337},
  year={2021},
  publisher={Springer}
}

@inproceedings{deitke2023objaverse,
  title={Objaverse: A universe of annotated 3d objects},
  author={Deitke, Matt and Schwenk, Dustin and Salvador, Jordi and Weihs, Luca and Michel, Oscar and VanderBilt, Eli and Schmidt, Ludwig and Ehsani, Kiana and Kembhavi, Aniruddha and Farhadi, Ali},
  booktitle={Proceedings of the IEEE/CVF conference on computer vision and pattern recognition},
  pages={13142--13153},
  year={2023}
}

@article{dielen2021learning,
  title={Learning Direction Fields for Quad Mesh Generation, Eurographics Symposium on Geometry Processing},
  author={Dielen, A and Lim, I and Lyon, M and Kobbelt, L},
  journal={Ed. K. Crane and J. Digne},
  volume={40},
  number={5},
  year={2021}
}

@article{deitke2023objaversexl,
  title={Objaverse-xl: A universe of 10m+ 3d objects},
  author={Deitke, Matt and Liu, Ruoshi and Wallingford, Matthew and Ngo, Huong and Michel, Oscar and Kusupati, Aditya and Fan, Alan and Laforte, Christian and Voleti, Vikram and Gadre, Samir Yitzhak and others},
  journal={Advances in Neural Information Processing Systems},
  volume={36},
  pages={35799--35813},
  year={2023}
}

@inproceedings{chen2025dora,
  title={Dora: Sampling and benchmarking for 3d shape variational auto-encoders},
  author={Chen, Rui and Zhang, Jianfeng and Liang, Yixun and Luo, Guan and Li, Weiyu and Liu, Jiarui and Li, Xiu and Long, Xiaoxiao and Feng, Jiashi and Tan, Ping},
  booktitle={Proceedings of the IEEE/CVF Conference on Computer Vision and Pattern Recognition},
  pages={16251--16261},
  year={2025}
}

@article{zhao2023michelangelo,
  title={Michelangelo: Conditional 3d shape generation based on shape-image-text aligned latent representation},
  author={Zhao, Zibo and Liu, Wen and Chen, Xin and Zeng, Xianfang and Wang, Rui and Cheng, Pei and Fu, Bin and Chen, Tao and Yu, Gang and Gao, Shenghua},
  journal={Advances in neural information processing systems},
  volume={36},
  pages={73969--73982},
  year={2023}
}

@article{zhang20233dshape2vecset,
  title={3dshape2vecset: A 3d shape representation for neural fields and generative diffusion models},
  author={Zhang, Biao and Tang, Jiapeng and Niessner, Matthias and Wonka, Peter},
  journal={ACM Transactions On Graphics (TOG)},
  volume={42},
  number={4},
  pages={1--16},
  year={2023},
  publisher={ACM New York, NY, USA}
}

@inproceedings{campen2017partitioning,
  title={Partitioning Surfaces into Quad Patches.},
  author={Campen, Marcel},
  booktitle={Eurographics (Tutorials)},
  year={2017}
}

@article{sumner2004deformation,
  title={Deformation transfer for triangle meshes},
  author={Sumner, Robert W and Popovi{\'c}, Jovan},
  journal={ACM Transactions on graphics (TOG)},
  volume={23},
  number={3},
  pages={399--405},
  year={2004},
  publisher={ACM New York, NY, USA}
}

@article{tao2025learning,
  title={Learning conjugate direction fields for planar quadrilateral mesh generation},
  author={Tao, Jiong and Yang, Yong-Liang and Deng, Bailin},
  journal={arXiv preprint arXiv:2511.11865},
  year={2025}
}

\end{document}